\documentclass[runningheads]{llncs}
\usepackage{times}
\usepackage{graphicx}
\usepackage{amssymb}
\usepackage{url,hyperref}
\usepackage{amsfonts}
\usepackage{bbold}
\usepackage{dsfont}
\usepackage{amsmath}
\usepackage{amssymb}
\usepackage{bbm}
\usepackage{CJKutf8}

\usepackage{tabularx}
\usepackage{makecell}
\usepackage{graphicx}
\usepackage{multirow}
\usepackage{ulem}

\begin{document}


\title{Adaptive Semi-Supervised Segmentation of Brain Vessels with Ambiguous Labels
}
%
%
\author{Fengming Lin\inst{1}\and
Yan Xia\inst{1} \and
Nishant Ravikumar\inst{1} \and
Qiongyao Liu\inst{1}\and
Michael MacRaild\inst{1}\and
Alejandro F Frangi\inst{2}
}
\authorrunning{Fengming Lin et al.}
%
\institute{University of Leeds, UK \and
University of Manchester, UK
}
\maketitle              
\vspace{-0.5cm} 
\begin{abstract}
Accurate segmentation of brain vessels is crucial for cerebrovascular disease diagnosis and treatment. However, existing methods face challenges in capturing small vessels and handling datasets that are partially or ambiguously annotated. In this paper, we propose an adaptive semi-supervised approach to address these challenges. Our approach incorporates innovative techniques including progressive semi-supervised learning, adaptative training strategy, and boundary enhancement. Experimental results on 3DRA datasets demonstrate the superiority of our method in terms of mesh-based segmentation metrics. By leveraging the partially and ambiguously labeled data, which only annotates the main vessels, our method achieves impressive segmentation performance on mislabeled fine vessels, showcasing its potential for clinical applications.


\end{abstract}

\vspace{-0.5cm} 
\section{Introduction}
Accurate segmentation of cerebral vessels is clinically significant as it provides crucial anatomical information for the diagnosis and assessment of cerebrovascular diseases \cite{ma2021exploring}. 
Furthermore, segmenting small vessels is essential as they play important roles in brain function and pathological processes. Accurate segmentation of small vessels provides comprehensive morphological information about the vascular network, facilitating patient-specific modeling of cerebral hemodynamics, which can be used to better understand pathologies, plan interventions, and design treatment devices
\cite{hilbert2020brave}\cite{aydin2021evaluation}\cite{dai2015new}\cite{ciecholewski2021computational}.


However, the task of accurate vessel segmentation is challenging due to several reasons. Firstly, the small proportion of vessels in brain tissue makes segmentation difficult, particularly for small arterioles \cite{fu2020rapid}\cite{law2007vessel}.
To address this, convolution-based methods \cite{isensee2021nnu} designed for medical imaging are enhanced in segmentation capability through experienced pre-processing and post-processing techniques. Further, transformer-based methods \cite{cao2022swin}\cite{chen2021transunet} have been proposed to leverage fully supervised learning to explore the features of small targets in-depth.
Secondly, 
clinical vessel annotations are focused only on regions surrounding pathologies such as aneurysms \cite{krings2011intracranial}\cite{samaniego2019vessel}, and only the main vessels are labeled, leaving out fine vessels. This ambiguously-labeled data negatively impacts the performance of fully supervised learning approaches. To overcome this limitation, semi-supervised learning methods with pseudo-labeling techniques \cite{chen2021semi}\cite{chatterjee2022ds6} have been proposed.
Thirdly, clinical images often exhibit high levels of noise \cite{benkner2010neurist}, and there are significant variations in pixel distribution across different imaging centers.
Traditional semi-supervised methods\cite{nie2018asdnet}\cite{chen2019multi}\cite{luo2021semi}\cite{jiao2022learning} using pseudo-labeling \cite{chen2021semi} tend to overly incentivize the confidence of the model in vessel segmentation, leading to excessive over-segmentation \cite{rizve2021defense}.

\begin{figure}[!ht]
\centerline{\includegraphics[width=\columnwidth]{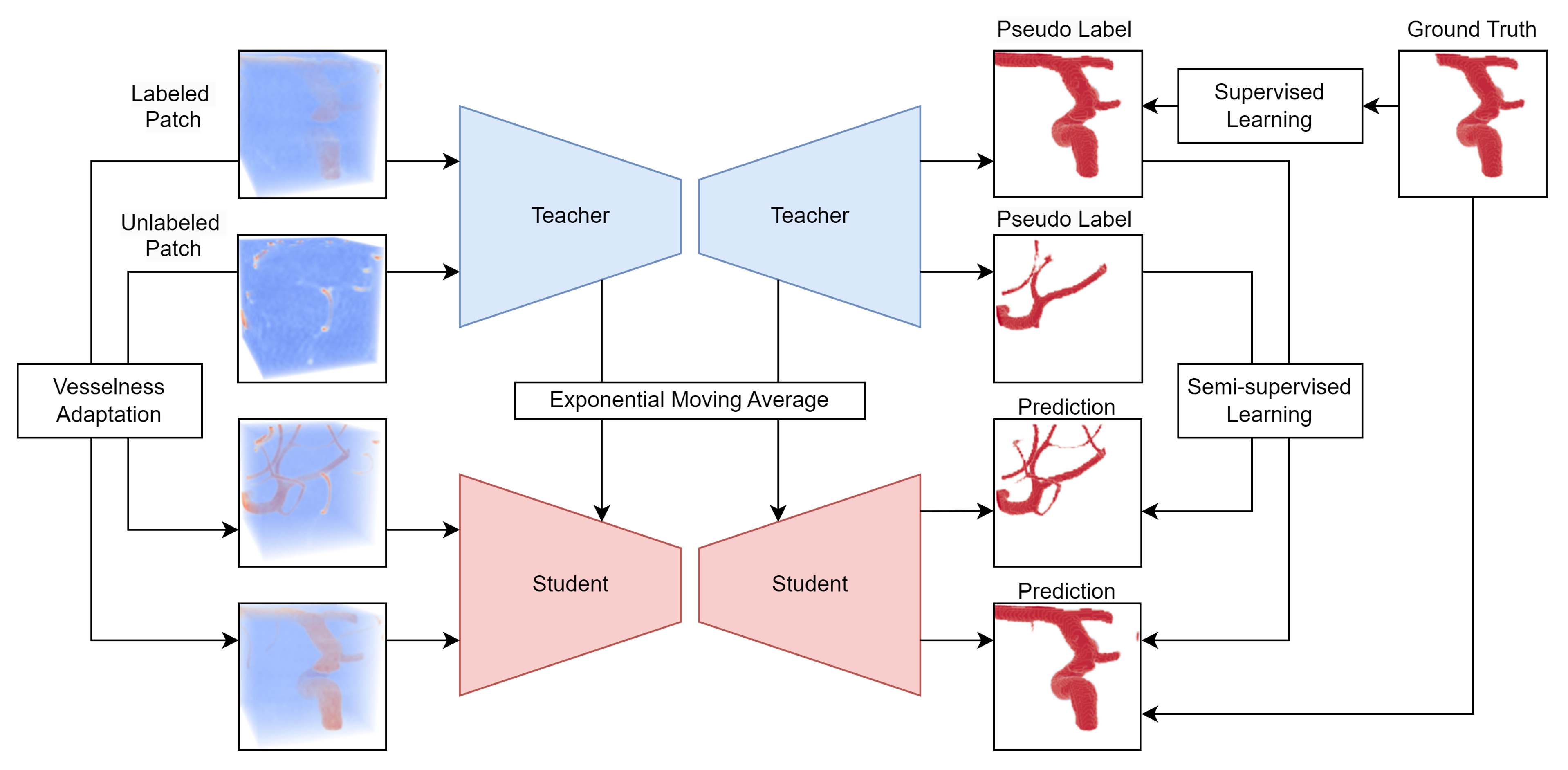}}
\caption{Schematic of the proposed adaptive semi-supervised model.}
\label{main01_01}
\end{figure}

Therefore, we propose the adaptive semi-supervised model in Fig.\ref{main01_01}, aiming to address the challenge of partially annotated intracranial vessel segmentation. The model employs a Teacher-Student structure, with the Swin-UNet \cite{cao2022swin} serving as the backbone network. We partition the partially annotated data into labeled patches and unlabeled patches, which are fed into the teacher and student networks, respectively. The teacher network learns vessel weight knowledge from the labeled patches and shares it with the student network. Additionally, the teacher network's output is used as refined pseudo-labels for further learning by the student network.
The key innovations are as follows:
\begin{itemize}
    \item[$\bullet$] We introduce the adaptive semi-supervised model, utilizing a progressive semi-supervised learning strategy. Ground truth is used to teach the teacher network, and the teacher network in turn instructs the student network, leading to incremental improvements in segmentation performance.
    \item[$\bullet$] We propose addressing the challenges associated with semi-supervised learning through unsupervised domain adaptation techniques. This enables the adaptation of knowledge from labeled patches to unlabeled patches without any domain shift.
    \item[$\bullet$] We introduce the Fourier high-frequency boundary loss. Except for Dice and Cross-Entropy loss, we extract the high-frequency boundary features using the Fourier transform and calculate their mean squared error. 
    \item[$\bullet$] We introduced a data augmentation technique called adaptive histogram attention (AHA) to address the variations in pixel distribution within clinical data. 
    AHA enables the model to better focus on discriminating between other brain tissues and vessels, facilitating the extraction of vessel structural features. 
\end{itemize}

\vspace{-0.5cm} 
\vspace{-0.5cm} 
\section{Methodology}
\vspace{-0.1cm} 
\subsection{Preprocessing}
\vspace{-0.1cm} 
The purpose of preprocessing is not only to remove noise but also to facilitate the model in extracting size, structural, and generalization features specific to the vessels. 

Resolution standardization:  To tackle the resolution inconsistencies in clinical data, all data is standardized to a spacing of 0.35mm/pixel. This allows the model to learn the size/shape features in cases with initially different resolutions.

Adaptive Histogram Attention: 
The distribution patterns in 3DRA data histograms show that vessels are typically in the higher pixel value range, brain tissues in the middle, and backgrounds in the lower range. While deep learning models can easily distinguish between the background and vessels, they may struggle with distinguishing between brain tissues and vessels, leading to over-segmentation. AHA tackles this by identifying areas of abrupt shifts in the histogram, using them for normalization, effectively eliminating the background from the histogram, thereby enabling the model to focus on distinguishing between vessels and other brain tissues. This method emphasizes the extraction of structural features rather than mere threshold-based features.


Patch Grouping: 
We extract overlapped 3D patches from both the annotated and unannotated regions. 
These two groups of patches are subsequently fed into the teacher and student networks, respectively. This approach
enables the model to learn local generalized vessel features rather than specific fitting features of individual cases.
\vspace{-0.4cm} 
\subsection{Problem Formulation}
\vspace{-0.1cm} 
In this paper, we have datasets sampled from two groups. The labeled group contains labeled patch $\mathcal{D}_{l}=\left\{\left(x_{i}^{l}, y_{i}^{l}\right)\right\}_{i=1}^{N_{l}}$, and the unlabled group contains unlabeled patch $\mathcal{D}_{u}=\left\{\left(x_{i}^{u}\right)\right\}_{i=1}^{N_{u}}$. We use adaptive histogram attention to get labeled and unlabeled vessel-like patch $\mathcal{D}_{\hat{l}}=\left\{\left(\hat{x}_{i}^{l} \right)\right\}_{i=1}^{N_{l}}$ and $\mathcal{D}_{\hat{u}}=\left\{\left(\hat{x}_{i}^{u} \right)\right\}_{i=1}^{N_{u}}$.
Our model consists of teacher and student networks. We update the weights in the student network (encoder $F_{s}$, decoder $G_{s}$) as an exponential moving average (EMA) of weights in the teacher network (encoder $F_{t}$, decoder $G_{t}$) to ensemble the information in different training steps. 
The prediction of teacher network on labeled and unlabeled patch are denoted as $p_{i}^{l} = G_{t}\left(F_{t}\left(x_{i}^{l}\right)\right)$ and ${p}_{i}^{u} = G_{t}\left(F_{t}\left({x}_{i}^{u}\right)\right)$. We also denote the prediction of the student network on
labeled and unlabeled vessel-like patch as $\hat{p}_{i}^{l} = G_{s}\left(F_{s}\left(\hat{x}_{i}^{l}\right)\right)$ and $\hat{p}_{i}^{u} = G_{s}\left(F_{s}\left(\hat{x}_{i}^{u}\right)\right)$.
Our goal is to learn a task-specific student network using ${F}_{s}$ and ${G}_{s}$ to accurately predict labels on test data from the unlabeled patches.
\vspace{-0.4cm} 
\subsection{Supervised Learning}
\vspace{-0.1cm} 
In the teacher network, labeled patches $\mathcal{D}_{l}$ are passed through the CNN-based feature extractor $F_{t}$, which are then passed through the task-specific segmentation generator $G_{t}$ to minimize the supervised loss $\mathcal{L}_{sup}$ which includes cross-entropy $\mathcal{L}_{CE}$, Dice similarity coefficient loss $\mathcal{L}_{DSC}$, and boundary loss $\mathcal{L}_{B}$.

\begin{equation}
\mathcal{L}_{full\underline{\hspace{0.3em}}sup} = \mathcal{L}_{CE} + \mathcal{L}_{DSC} + \mathcal{L}_{B}
\end{equation}
\begin{equation}
\begin{split}
\mathcal{L}_{CE} &= - \frac{1}{N_{l}} \sum_{i=1}^{N_{l}} y_{i}^{l}\log \left(p_{i}^{l}\right) 
\end{split}
\end{equation}
\begin{equation}
\begin{split}
\mathcal{L}_{DSC} 
&= \frac{1}{N_{l}} \sum_{i=1}^{N_{l}} \left ( 1 -  \frac{2\left | p_{i}^{l} \cap y_{i}^{l} \right | }{\left |  p_{i}^{l} \right | + \left | y_{i}^{l} \right |} \right ) 
\end{split}
\end{equation}
\begin{equation}
\begin{split}
\mathcal{L}_{B} &= \frac{1}{N_{l}}\sum_{i=1}^{N_{l}} \left(   H \left( p_{i}^{l} \right) -  H \left( y_{i}^{l} \right)  \right)  ^{2}
\end{split}
\end{equation}
\begin{equation}
\begin{split}
H \left( p_{i}^{l} \right) = \mathcal{F}^{-1}  \left( \mathcal{F} \left( p_{i}^{l} \right) \cdot \mathbb{1}_{mask}  \right)
\end{split}
\end{equation}

where $\mathcal{F}$ and $\mathcal{F}^{-1}$ are the Fourier transform \cite{cochran1967fast} and the Fourier inverse transform respectively. The mask $\mathbb{1}_{ mask }$ with value one in the middle and value zero on the edge has the same shape as $p_{i}^{l}$.
\vspace{-0.3cm} 
\subsection{Semi-supervised Learning}
\vspace{-0.1cm} 
To perform alignment at the instance level, the adaptive vessel-like labeled and unlabeled patches are passed through the teacher network to get segmentation prediction ${p}_{i}^{l}$ and ${p}_{i}^{u}$.
Meanwhile, we generate the adaptive vessel-like labeled and unlabeled patches, and they are passed through the student network to get segmentation prediction $\hat{p}_{i}^{l}$ and $\hat{p}_{i}^{u}$. 
Next, we employ the mean square error (MSE) \cite{willmott2005advantages} and cosine similarity \cite{lahitani2016cosine} as defined in Eq \ref{LMSE} and Eq \ref{LCS}  to reduce the discrepancy between the two predictions and thus increase the vessel invariance of the student model.
\begin{equation}
\mathcal{L}_{semi\underline{\hspace{0.3em}}sup} =  \mathcal{L}_{mse} +  \mathcal{L}_{sim}
\end{equation}
\begin{equation}
\label{LMSE}
\mathcal{L}_{mse} = \frac{1}{N_{l}}\sum_{i=1}^{N_{l}}\left(p_{i}^{l}-\hat{p}_{i}^{l}\right)^2 + \frac{1}{N_{u}}\sum_{i=1}^{N_{u}}\left(\hat{p}_{i}^{u}-p_{i}^{u}\right)^2
\end{equation}
\begin{equation}
\label{LCS}
\mathcal{L}_{sim}=\frac{1}{N_{l}} \sum_{i=1}^{N_{l}} h\left(p_{i}^{l}, \hat{p}_{i}^{l}\right) + \frac{1}{N_{u}} \sum_{i=1}^{N_{u}} h\left(\hat{p}_{i}^{u}, p_{i}^{u}\right)
\end{equation}
\begin{equation}
h(u, v)=\exp \left(\frac{u^{T} v}{\|u\|_{2}\|v\|_{2}}\right)
\end{equation}
The weight ratio between fully supervised and semi-supervised losses is 4:1. We prioritize the fully supervised loss to ensure training robustness and prevent the network from becoming overly confident and introducing noise during the initialization stage.
\section{Experiments and Results}
\vspace{-0.3cm} 
\subsection{Datasets}
\begin{figure}[!ht]
\centerline{\includegraphics[width=\columnwidth]{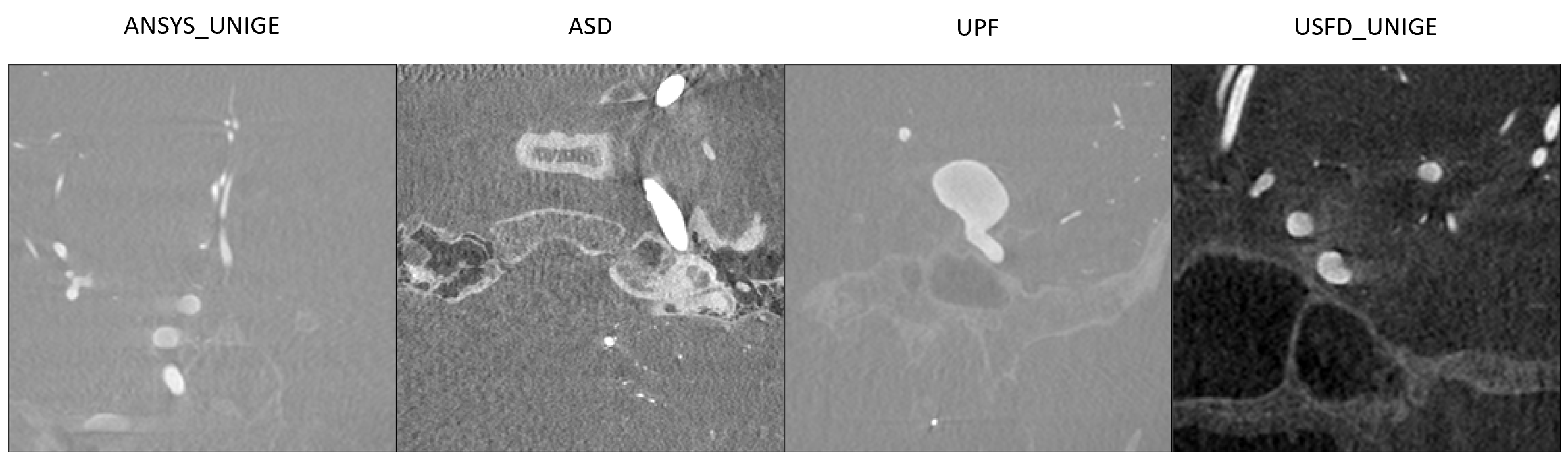}}
\caption{Examples of 3DRA images in grayscale collected from AneurIST dataset: 2D visualization of data from four different sources showed great differences in pixel distribution and noise levels.}
\label{fig02}
\end{figure}

\vspace{-0.1cm} 
In our experiments, we utilized 3D rotational angiography (3DRA) modality dataset Aneurist \cite{benkner2010neurist}, which comprises 223 partially annotated 3D brain vessel images. These images were acquired from four different centers using different scanners and imaging protocols. As shown in Fig.~\ref{fig02}, there are significant variations in image appearance and resolution across the data from different centers. 
we trained our models using full-size images and partially and ambiguously annotated labels. Due to the incomplete annotations, quantitative analysis was performed within the bounding box of annotated regions, while qualitative analysis was conducted across the entire image. 
\vspace{-0.5cm} 
\vspace{-1cm} 
\vspace{-0.3cm} 
\subsection{Experimental Setup}
\vspace{-0.1cm} 
The experiments were conducted using a high-performance computing setup. We utilized an NVIDIA GeForce RTX 3090 GPU with 24GB of VRAM. The experimental system was equipped with a high-capacity RAM of 128GB, enabling the handling of large datasets and memory-intensive tasks.

Our proposed adaptive semi-supervised model was implemented based on the Swin-UNet architecture \cite{cao2022swin}, serving as the backbone of both teacher and student networks. 
During training, we employed a batch size of 1 and utilized patch-based learning with a patch size of [128, 128, 128]. The models were trained for 100 epochs, during which the optimization was performed using the Adam optimizer \cite{kingma2014adam}. We employed data augmentation techniques, such as random rotations and flip, to enhance model generalization. The learning rate was initially set to 0.001, and a learning rate decay strategy was applied, reducing the learning rate by a factor of 0.1 every ten epochs. 
The parameters of the teacher network are updated normally, and the parameters of the student network are updated according to the exponential moving average (EMA) \cite{wang2020noise}. To achieve the effect of the EMA during the training process, we employed the no gradient decorator to ensure that gradient calculations are not performed during the EMA process. Prior to optimizing the parameters, we updated the parameters of the teacher network by invoking the EMA function in Eq \ref{eq13}. 
In this function, the weight factor $decay$ is calculated based on the iteration count and the initial decay rate of 0.999 in Eq.\ref{eq14}. The student network parameters are updated by applying the weight factor $decay$, thereby incorporating the knowledge learned by the teacher network gradually.
\begin{equation}
\mathcal{W}_{\text{stu}} = decay \times \mathcal{W}_{\text{stu}} + (1 - decay) \times \mathcal{W}_{\text{tea}}
\label{eq13}
\end{equation}
\vspace{-0.2cm} 
\begin{equation}
decay = \min \left(1 - \frac{1}{{iteration \times 10 + 1}}, decay\right)
\label{eq14}
\end{equation}
\vspace{-0.2cm} 

The image data from these databases were split on a patient-wise basis into training, validation, and test sets using a ratio of 7:1:2, respectively. To ensure a thorough evaluation of the segmentation performance, we performed five-fold cross-validation experiments, with the test sets in different cross-validation folds covering the entire dataset.

\subsection{Evaluation Metrics}

We utilized six evaluation metrics to assess the segmentation performance of our method:
Dice similarity coefficient (DSC): Measures the overlap between predicted and ground truth segmentations.
Sensitivity: Calculates the proportion of correctly identified positive instances.
Precision: Quantifies the accuracy of positive predictions.
Specificity: Measures the ability to correctly identify negative instances.
Jaccard index (Jac): Evaluates the overall agreement between predicted and ground truth segmentations.
Volume similarity (VS): Measures the similarity of segmented volume with the ground truth.

However, due to the ambiguous annotation of the dataset, most of the fine vessels are not labeled. This can lead to situations where segmentation results with higher accuracy actually have lower DSC. To provide a more comprehensive evaluation of segmentation performance, we employed surface-to-surface distance error (Surface Error) metrics to measure segmentation accuracy based on mesh representations. Furthermore, in our qualitative analysis, we employ visualization techniques to further evaluate the segmentation results, including the degree of over-segmentation and the accuracy of fine vessel segmentation.

The surface error metrics estimate the error between the ground-truth surfaces $S$, and the segmentation prediction surfaces $S{}'$. The distance between a point $p_{i}$ on surface $S$ and the nearest point on surface $S{}'$ is given by the minimum of the Euclidean norm. And we compare the similarity between the prediction and ground-truth by generating surface mesh-based representations of these structures from their corresponding masks in Eq.\ref{eqse1}. Doing this for all $N$ points in the ground-truth surface $S$ gives the average surface-to-surface distance error in Eq.\ref{eq8}. The p-value is calculated based on surface error.
\begin{equation}
d(p_{i},S{}')=\min_{p{}'\in S{}'}\left \| p_{i}-p{}' \right \|_{2}
\label{eqse1}
\end{equation}
\vspace{-0.2cm} 
\begin{equation}
d(S,S{}') = \frac{1}{N}\sum_{N}^{i}d(p_{i},S{}')
\label{eq8}
\end{equation}
\subsection{Qualitative  Results and Analysis}
\begin{figure}[!ht]
\centerline{\includegraphics[width=\columnwidth]{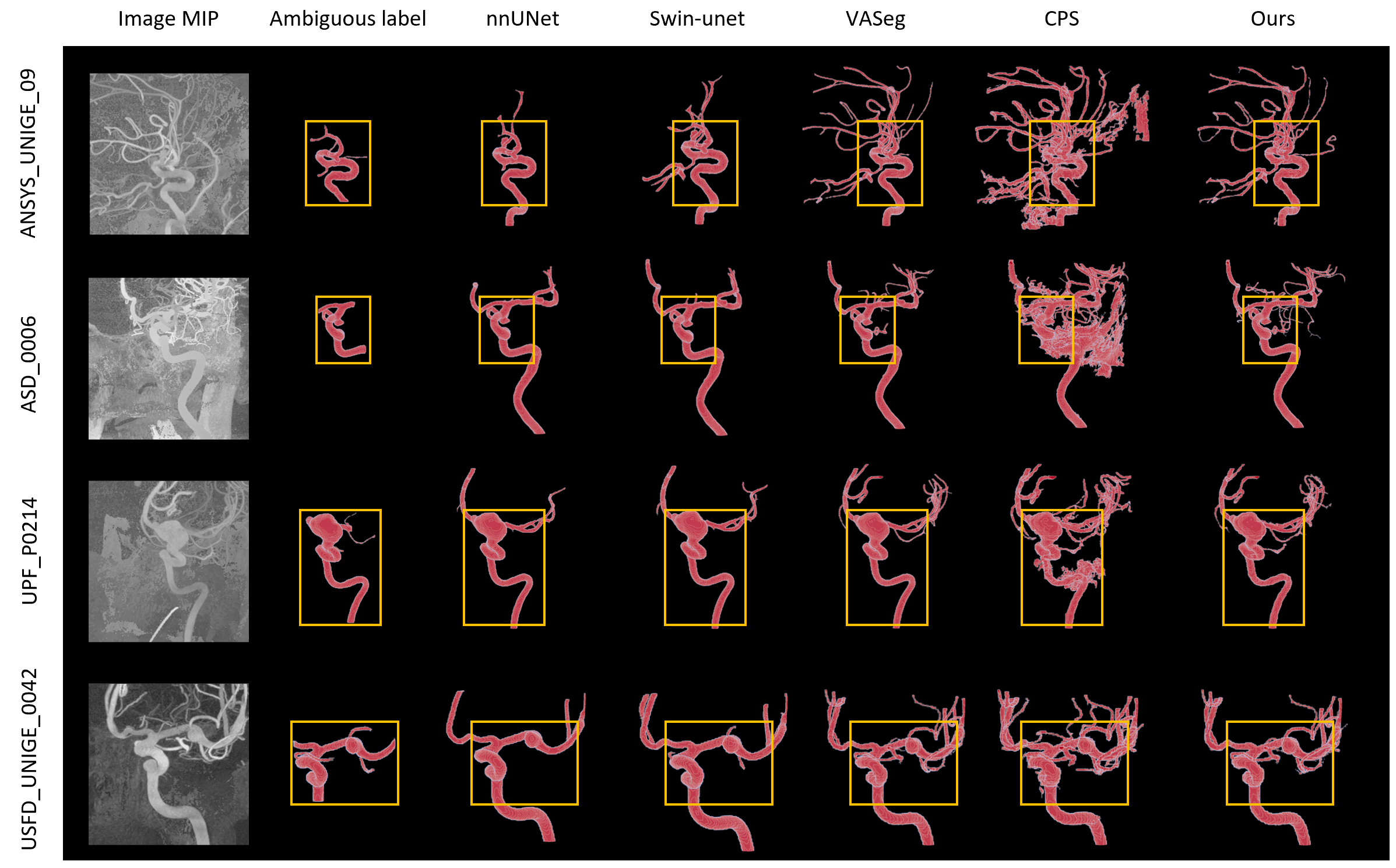}}
\caption{Comparison with State-of-the-Art Methods on four different data sources. The yellow box is the golden standard area where all quantitative evaluations are carried out.}
\label{fig03}
\end{figure}

In Fig. \ref{fig03}, we compared the proposed method with state-of-the-art approaches on four different data sources. Our method demonstrated superior performance in segmenting fine vessels without introducing excessive over-segmentation noise, especially in datasets with high levels of noise, such as ANSYS, ASD, and UPF. Notably, the nnUNet \cite{isensee2021nnu}
was greatly affected by ambiguous labels and could only segment major vessels. The Swin-UNet \cite{cao2022swin}, utilizing the swin-transformer structure for feature extraction, outperformed convolution in nnUNet by extracting a larger number of vessel branches. VASeg \cite{lin2023high}, employing majority voting and thresholding techniques, achieved a better recovery of fine vessels. CPS \cite{chen2021semi}, due to the utilization of semi-supervised cross pseudo-supervision,
exhibited increased segmentation uncertainty and introduced excessive noise when handling datasets with higher noise levels. 
In contrast, our method showcased the ability to segment a significant number of fine vessels while maintaining robustness and avoiding the introduction of excessive noise.
\vspace{-0.5cm} 

\subsection{Quantitative Results and Analysis}
Due to the uncertain and ambiguous nature of our dataset annotations, where only the main vessels near the aneurysm are labeled, we utilized mesh-based evaluation metrics as our primary performance measure, while pixel-based evaluation metrics such as the DSC were used as supplementary reference metrics.

In Table. \ref{table1}, we compared our method with several approaches, including the convolution-based fully supervised method nnUNet, the transformer-based fully supervised method Swin-UNet, the VASet method that addresses ambiguous label issues through preprocessing and postprocessing, and the traditional semi-supervised method CPS using cross pseudo-supervision. By comparing the surface error, we found that our method achieved the highest accuracy in vessel surface segmentation, with an average mesh error of 0.20mm (0.35mm/pixel).
Additionally, it is worth noting that our method achieves high sensitivity, second only to CPS. This indicates that our method successfully identifies a larger proportion of positive instances, meaning that it effectively captures the majority of the annotated vessels. 

\begin{table*}[!htbp]
\centering
\caption{Compare with state-of-the-art on whole Aneurist dataset. 
The mesh-based surface error serves as the primary evaluation metric, while pixel-based metrics such as the DSC are used as supplementary evaluation criteria due to incomplete annotation. Quantitative analysis was performed within the annotated regions instead of the full image.
}\label{table1}
\resizebox{\textwidth}{!}{
\begin{tabular}{c|c|c|c|c|c}
\hline
\hline
Methods&nnUNet&Swin-unet&VASeg&CPS&Ours\\
\hline
Sensitivity
&\makecell[l]{0.9196 ± 0.0637}
&\makecell[c]{0.8967 ± 0.1219}
&\makecell[l]{0.9572 ± 0.0510}
&\makecell[l]{0.9872 ± 0.0189}
&\makecell[l]{\textbf{0.9793 ± }\textbf{0.0183}
}\\
Precision
&\makecell[l]{0.9300 ± 0.0345}
&\makecell[c]{0.8689 ± 0.0487}
&\makecell[l]{0.8792 ± 0.0761}
&\makecell[l]{0.6636 ± 0.1238}
&\makecell[l]{\textbf{0.8018 ± }\textbf{0.0877}
}\\
Specificity
&\makecell[l]{0.9956 ± 0.0028}
&\makecell[c]{0.9932 ± 0.0032}
&\makecell[l]{0.9934 ± 0.0044}
&\makecell[l]{0.9740 ± 0.0129}
&\makecell[l]{\textbf{0.9881 ± }\textbf{0.0056}
}\\

Jac
&\makecell[l]{0.8605 ± 0.0680}
&\makecell[c]{0.7889 ± 0.1091}
&\makecell[l]{0.8440 ± 0.0763}
&\makecell[l]{0.6572 ± 0.1209}
&\makecell[l]{\textbf{0.7873 ± } \textbf{0.0817}}\\
VS
&\makecell[l]{0.9728 ± 0.0263}
&\makecell[c]{0.9424 ± 0.0649}
&\makecell[l]{0.9450 ± 0.0509}
&\makecell[l]{0.7968 ± 0.0996}
&\makecell[l]{\textbf{0.8977 ± } \textbf{0.0602}}\\
DSC
&\makecell[l]{0.9236 ± 0.0408}
&\makecell[c]{0.8772 ± 0.0787}
&\makecell[l]{0.9134 ± 0.0479}
&\makecell[l]{0.7862 ± 0.0960}
&\makecell[l]{\textbf{0.8786 ± }\textbf{0.0536}}\\
\hline
Surface Error
&\makecell[l]{0.8903 ± 1.2450}
&\makecell[c]{0.6801 ± 1.6093}
&\makecell[l]{0.2586 ± 0.3066}
&\makecell[l]{0.3068 ± 0.1210}
&\makecell[l]{\textbf{0.2075 ±} \textbf{0.0640}}\\
\hline
\makecell[c]{p-value}&
\makecell[c]{\textless0.05}&
\makecell[c]{\textless0.05}&
\makecell[c]{\textless0.05}&
\makecell[c]{\textless0.05}&
\makecell[c]{/}
\\
\hline
\end{tabular}
}
\end{table*}

Table.\ref{table2} presents the results of the ablation study conducted to analyze the impact of different components in our method. We used the fully supervised Swin-UNet as the baseline, trained solely using the teacher network with Dice Cross Entropy loss.
In the second set of experiments, we introduced the Fourier boundary loss to the fully supervised loss. The inclusion of this loss led to a noticeable improvement in surface error, indicating enhanced boundary delineation.
In the third set of experiments, we employed vessel adaptation by feeding the data into the student network. We also incorporated the semi-supervised loss to train the student network. The results showed a significant increase in the number of predicted vessels, as evidenced by the improved sensitivity. Additionally, the surface error achieved a level of 0.20mm, indicating precise vessel segmentation at the boundary.

\begin{table*}[!htbp]
\centering
\caption{Ablation study.}\label{table2}
\begin{tabular}{c|c|c|c}
\hline
\hline
\makecell[c]{Modules}
&\makecell[c]{Supervised Loss\\(Dice + CE)}
&\makecell[c]{Supervised Loss\\Boundary Loss}
&\makecell[c]{Supervised Loss\\Boundary Loss\\Semi-supervised Loss}
\\
\hline
Sensitivity
&\makecell[c]{0.8967 ± 0.1219}
&\makecell[c]{0.9236 ± 0.0600}
&\makecell[c]{\textbf{0.9793 ± }\textbf{0.0183}}\\

Precision
&\makecell[c]{0.8689 ± 0.0487}
&\makecell[c]{0.9116 ± 0.0577}
&\makecell[c]{\textbf{0.8018 ± }\textbf{0.0877}}\\

Specificity
&\makecell[c]{0.9932 ± 0.0032}
&\makecell[c]{0.9954 ± 0.0035}
&\makecell[c]{\textbf{0.9881 ± }\textbf{0.0056}}\\

Jac
&\makecell[c]{0.7889 ± 0.1091}
&\makecell[c]{0.8456 ± 0.0652}
&\makecell[c]{\textbf{0.7873 ± }\textbf{0.0817}}\\

VS
&\makecell[c]{0.9424 ± 0.0649}
&\makecell[c]{0.9636 ± 0.0440}
&\makecell[c]{\textbf{0.8977 ± }\textbf{0.0602}}\\

DSC
&\makecell[c]{0.8772 ± 0.0787}
&\makecell[c]{0.9149 ± 0.0417}
&\makecell[c]{\textbf{0.8786 ± }\textbf{0.0536}}\\
\hline
Surface Error
&\makecell[c]{0.6801 ± 1.6093}
&\makecell[c]{0.3483 ± 0.3111}
&\makecell[c]{\textbf{0.2075 ±} \textbf{0.0640}}\\
\hline
\makecell[c]{p-value}&
\makecell[c]{\textless0.05}&
\makecell[c]{\textless0.05}&
\makecell[c]{/}
\\
\hline
\hline
\end{tabular}
\end{table*}
\vspace{-1cm} 
\section{Conclusion}
In summary, our semi-supervised model brings forward innovative techniques for cerebral vessel segmentation. With semi-supervised learning and domain adaptation-like strategies, Fourier high-frequency boundary loss, and adaptive histogram attention, we achieve better segmentation accuracy and robustness on whole vessels, paving the way for clinical uses such as treatment planning. However, our model might underperform on lower noise datasets due to our focus on robustness. Future research will explore contrastive learning to improve performance on low-noise datasets.


\end{document}